\documentclass[10pt,letterpaper,twocolumn]{article} 

\usepackage{graphicx}
\usepackage{ol}
\usepackage{amsmath}

\begin{document}

\twocolumn[ 

\title{Impedance matching in photonic crystal microcavities for Second Harmonic Generation }

\author{Andrea Di Falco}
\address{%
N\it{oo}EL - Nonlinear Optics and OptoElectronics Laboratory, \\
Department of Electronic Engineering and INFM-CNISM,\\
University ``Roma Tre'',
Via della Vasca Navale, 84 - 00146 - Rome, Italy 
}%
\author{Claudio Conti}
\address{%
Research Center ``Enrico Fermi'', Via Panisperna 89/A - Rome, Italy\\
Research Center Soft INFM-CNR, University ``La Sapienza'', Piazzale Aldo Moro, 2 - 00185 - Rome, Italy
}%
\author{Gaetano Assanto}%
\address{%
N\it{oo}EL - Nonlinear Optics and OptoElectronics Laboratory\\
Department of Electronic Engineering and INFM-CNISM,\\
University ``Roma Tre'',
Via della Vasca Navale, 84 - 00146 - Rome, Italy 
}%

\date{\today}

\begin{abstract}
By numerically integrating the three-dimensional Maxwell equations in the time domain with reference to a dispersive 
quadratically nonlinear material, we study second harmonic generation in planar 
photonic crystal microresonators. The proposed scheme allows efficient coupling of the pump 
radiation to the defect resonant mode. The out-coupled generated second harmonic is maximized by impedance matching the photonic crystal cavity to the output waveguide.
\end{abstract}


 ] 

\maketitle

\bigskip

In the early sixties the very first experiments in nonlinear optics were Franken’s investigations 
of traveling-wave second harmonic generation.\cite{Franken} It took a number of years to realize 
that guided-wave optics and resonators could dramatically enhance phenomena requiring high intensities and tuning of critical parameters. \cite {Armstrong,Ashkin,Sohler}
Since then, both the advent of nano-optics and the technological advances have made new solutions available. Among them, microresonators can certainly be considered among the best candidates for nonlinear optics and frequency generations. \cite{Berger,Xu,Mookherjea,Spillane}
In the past few years, photonic crystal (PC) microcavities, i. e. periodic (bandgap) structures hosting a resonant defect, have attracted attention for some of their unique properties. \cite{Joannopulos_rev,Mingaleev}. 
In particular, in PC microresonators it is possible to obtain extremely high quality factors (Q) in reduced volumes while tailoring their dispersive features, 
\cite{Vuckovic,Srinivasan} characteristics which can be exploited to achieve efficient frequency generation with various schemes \cite{DiFalco_ol,Martorell,Mondia,Ishihara}.
The advantages inherent to a high confinement, however, are partially counterbalanced by a difficult energy coupling 
into the resonators: well isolated resonant states, in fact, correspond to large (external) quality factors of the cavity. \cite{HausBook}

To this extent, impedance matching has been proposed based on a properly designed coupling to/from the microcavities. \cite{Conti,Jugessur}
In this Letter we propose and investigate an efficient outsourcing scheme to maximize frequency doubling from a PC defect. 
Resorting to a second-order nonlinearity in a large Q photonic crystal microcavity, an optical pump is up-converted to a resonant cross-polarized harmonic signal. Since the device is designed to be nearly transparent to the pump wavelength, input coupling losses are minimized while the out-coupled second harmonic can be maximized by impedance matching.\\
\begin{figure}
\centerline{\includegraphics[width=8.3cm]{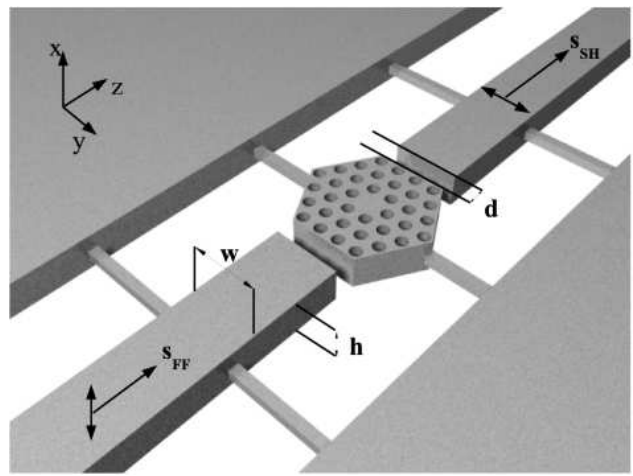}}
\caption{Artist's sketch of the structure.
\label{figure1}}
\end{figure}
The temporal evolution of the resonant mode at the second-harmonic (SH) can be described and related to the cavity parameters by coupled mode theory in the time domain (CMT-TD): \cite{HausBook}\begin{equation}
\label{CMT1}
\frac {da_{_{SH}}} {dt}=-i \omega_{_{SH}} a_{_{SH}}-\frac{a_{_{SH}}} {\tau_{_{SH}}}+\kappa s_{_{FF}}^2
\end{equation}
where $a_{_{SH}}$ is the mode amplitude at the SH frequency $\omega_{_{SH}}$, $\tau_{_{SH}}$ 
takes into account both internal ($1/\tau_o$) and external ($1/\tau_e$) losses  ($1/\tau_{_{SH}}=1/\tau_o+1/\tau_e$), $\kappa$ is the (three-dimensional)
overlap integral between the susceptibility tensor and the electric fields at fundamental (FF) and SH frequencies, $s_{_{FF}}$ is 
the amplitude of the FF guided mode propagating through the cavity via input and output waveguides. The latter amplitude acts as a source for SH generation in the resonator. A pictorial sketch of the PC microcavity with input and output guides is shown in Fig. 1.\\
Taking $s_{_{FF}}=\xi_{_{FF}}exp[-i \omega_{_{FF}} t]$ and $a_{_{SH}}=A_{_{SH}}exp[-i \omega_{_{SH}} t]$, 
with $\omega_{_{SH}}=2\omega_{_{FF}}$ yields the steady state solution of (1)
\begin{equation}
\label{CMT2}
A_{_{SH}} =\tau_{_{SH}}\kappa s_{_{FF}}^2.
\end{equation}
The out-coupled SH radiation is linked to the external losses by:
\begin{equation}
\label{CMT3}
s_{_{SH}} =\sqrt{\frac {2}{\tau_e}} A_{_{SH}}=\sqrt{\frac {2}{\tau_e}}\tau_{_{SH}}\kappa s_{_{FF}}^2.
\end{equation}
From the last expression, the normalized efficiency is then:
\begin{equation}
\label{CMT4}
\eta=\frac {|s_{_{SH}}|^2}{|s_{_{FF}}|^4} =\frac {2 \tau_e}{(1+\frac {\tau_e} {\tau_o})^2}|\kappa|^2= Q_o \frac {4} {\omega_{_{SH}}} \frac { Q_e/Q_o}{(1+\frac {Q_e} {Q_o})^2}|\kappa|^2,
\end{equation}
On the RHS of (4) 
$\tau_e$ and $\tau_o$ have been replaced by 
internal $Q_o$ and external $Q_e$ quality factors of the cavity, respectively, being 
$1/Q=2/(\tau \omega_{_{SH}})$. For a constant $\kappa$ in (4), the SHG efficiency is linearly dependent on $Q_o$ and 
maximized when $Q_e/Q_o=1$. Otherwise stated, once the cavity geometry is defined (i.e. given a certain $Q_o$) and for small changes in the overlap integral $\kappa$, the out-coupled SH can be optimized by engineering the factor $Q_e$.
\begin{figure}
\centerline{\includegraphics[width=8.3cm]{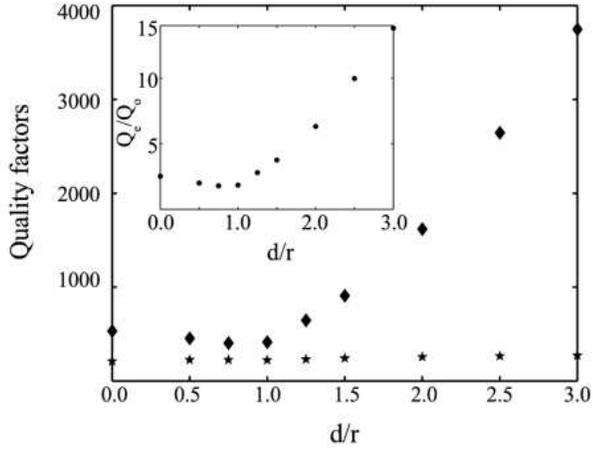}}
\caption{Quality factors $Q_e$ (diamonds) and $Q_o$ (stars) versus spacing $d$ (in units of $r$). The inset shows the ratio $Q_e/Q_o$ versus $d/r$.
\label{figure2}}
\end{figure}
In order to demonstrate the validity of the insight above, we need to take into account all the loss mechanisms 
(radiation, absorption and coupling), material and PC dispersion and nonlinear response. 
To this extent we employed a finite difference time domain (FDTD) code to study the nonlinear process without approximations.\cite{Taflove,DiFalco_apb} 
In the code, Maxwell equations in vacuum are coupled to the material polarization: 
\begin{equation}
\label{Maxwell}
\begin{array}{l}
\nabla\times{\bf E}=-\mu_0 \frac{\partial {\bf H}}{\partial t},~~ 
\nabla\times{\bf H}=\epsilon_0 \frac{\partial {\bf E}}{\partial t}+\frac{\partial {\bf P}}{\partial t}, \\
\\
\frac {\partial^2{\bf P}}{\partial{t^2}}+2\gamma_{0}\frac{\partial{{\bf P}}}{\partial{t}}+\omega_{R}^{2}{\bf P}+\underline{\underline{\bf D}}:{\bf PP} =\epsilon_{0}(\epsilon_{s}-1)\hat{\omega}_{R}^2 {\bf E}.
\end{array}
\end{equation}
The last equation describes a dipole oscillation with an anharmonic term accounting for 
the material quadratic response. 
Its polarization $\bf{P}$ is due to a single-pole dispersion with a Lorentzian distribution 
centered in $\omega_R$.
We chose a loss coefficient $\gamma_{0}=3.8 \times 10^8 s^{-1}$, 
$\omega_{R}=1.1406 \times 10^{16} s^{-1}$, a static permittivity $\epsilon_{s}=11.7045$ and $\hat\omega_{R}=1.0995 \times 10^{15} s^{-1}$ to mimic the response of GaAs. 
The pertinent tensor \underline{\underline{\bf D}} values were selected after a 45 degrees rotation in the plane $\hat{zy}$ of GaAs crystallographic axes.
This permits to generate (quasi-) TE polarized second harmonic when a linearly polarized (quasi-) TM pump-mode is launched in the structure through the input channel.
\begin{figure}
\centerline{\includegraphics[width=8.3cm]{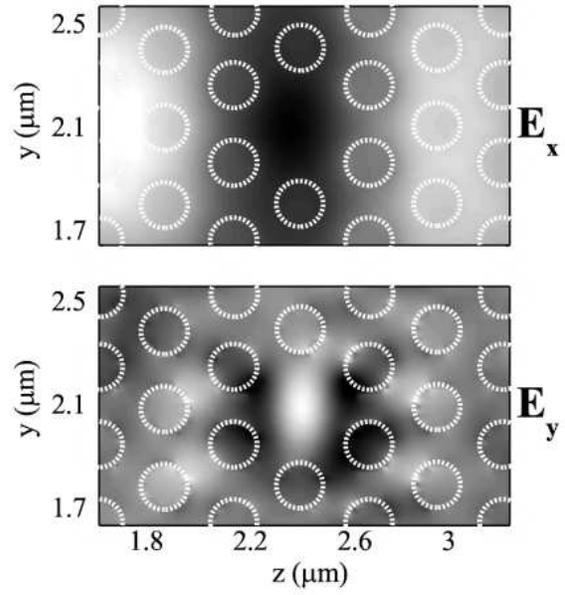}}
\caption{Snapshots of the electric field in the central portion of the PC microresonator. 
Top panel: FF x-component; bottom panel: SH y-component.}
\label{figure3}
\end{figure}
We investigated a microcavity consisting of a PC membrane with three concentric rings of holes with a central defect to support a resonant state at SH. The radius of the PC holes is $r=90 nm$ and the lattice constant $a=300 nm$.
Two suspended channel waveguides of width $w=1 \mu m$ and thickness $h=500 nm$ feed and out-couple the radiation to and from the microcavity, respectively. 
The separation $d$ can be varied in order to adjust the SH impedance of the waveguide to the resonant mode. For the sake of simplicity, we set the separation $d$ the same at both the input and the output.
The structure is designed to be resonant at 
$\lambda_{SH}=1.1370 \mu m $ for the TE polarization. 
The input waveguide couples a TM-polarized mode of amplitude $s_{_{FF}}$ at $\lambda_{FF}=2.2740 \mu m$ into the 
resonator, where it is up-converted to the cross-polarized SH 
mode of amplitude $a_{_{SH}}$ and out-coupled to the TE-mode $s_{_{SH}}$.
As $d$ was tuned, we verified that the resonant frequency did not change appreciably.
With the FDTD code, first we linearly characterized the defect state.
By evaluating the rate of energy decrease and the transmission $T$ at $\omega_{_{SH}}$,
$T=(1+Q_e/Q_o)^2$, we calculated both $Q_o$ and $Q_e$ for various $d$, 
according to:
\begin{equation}
\label{taus}
\begin{array}{l}
Q_e = \frac {Q_{_{SH}}Q_o} {Q_o-Q_{_{SH}}},~~~~
Q_o = Q_{_{SH}}[1+\frac {\sqrt{T}} {2(1-\sqrt{T})}]
\end{array}
\end{equation}
Fig. 2 displays the computed values: as expected, increasing the distance $d$ between the cavity and the waveguide(s) 
barely affects the internal (intrinsic) cavity losses, although an apparent increase takes place in the external quality 
factor owing to a lower coupling to and from the resonator.
The inset of Fig. 2 shows the calculated ratio between $Q_e$ and $Q_o$. On the basis of Eq. (\ref{CMT4}), the SH efficiency is maximum
when such ratio equals unity, i.e. for values of $d/r$ close to one (see inset).
We numerically launched a TM-polarized cw-like 
excitation at $\omega_{_{FF}}$ in the feeding waveguide.
The top panel of Fig. 3 is a snapshot of the TM component of the FF electric field, flowing from left to right through the PC and its defect. 
The TE-component of the generated electric field at SH is shown in the bottom panel of Fig. 3: the generated mode at $\omega_{_{SH}}$ clearly resonates in the central portion of the cavity.
To evaluate the SHG efficiency we computed the TE-polarized SH power in the output waveguide and 
scaled it to the square power of the FF radiation propagating through the PC.
The results are displayed in Fig. 4, which shows an apparent maximum in efficiency when 
the separation between waveguide(s) and PC cavity realizes the impedance matching, e.g. for $d/r=1$. Noteworthy, at variance with the simple CMT-TD, the latter ratio accounts for both the variations in $\kappa$ and the actual FF power coupled in the structure. 
\begin{figure}
\centerline{\includegraphics[width=8.3cm]{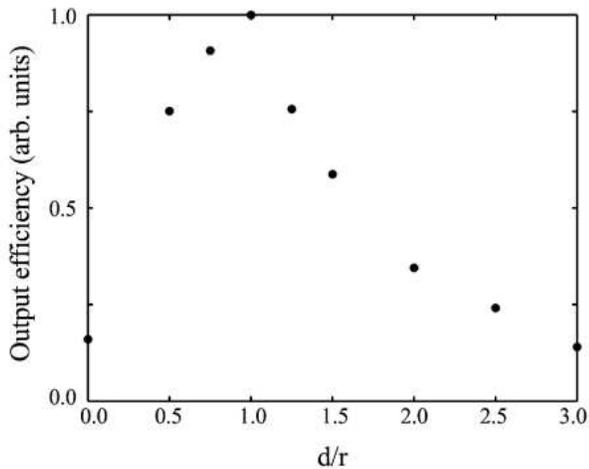}}
\caption{SH output efficiency versus separation $d$ (units of radius $r$).
\label{figure4}}
\end{figure}
For this case, we calculated an out-coupled SH-power $P_{_{SH}} = 0.22W$ when pumping with 
$P_{_{FF}}=1 KW$ through the input channel. Such value for $P_{_{SH}}$ needs be linked to the small quality 
factor $Q_o$ of the resonator (for computational reasons we analyzed a rather small PC) and to the non-optimized FF sourcing into the resonating mode at SH (i.e. a limited $\kappa$).
Both factors could be enhanced by a proper design.
In conclusion, we have numerically investigated by FDTD second harmonic generation in a photonic crystal microcavity, 
maximizing the out-coupled SH power by a simple impedance matching based on the separation between the PC and the output channel.
We anticipate that these results will promote design and realization of efficient integrated structures for all-optical signal handling.

The authors acknowledge the use of computer facilities at the Italian Consortium for Advanced Calculus (CINECA). A.D.F. is also with the Department of Electrical, Electronic and Telecommunication Engineering, University of Palermo, and thanks the Italian Electronic and Electrical Engineering Association (AEIT) for support.
\email{difalco@ele.uniroma3.it} 

\end{document}